\RequirePackage{amsmath}
\documentclass[epj]{svjour}

\usepackage{graphics}

\usepackage{amssymb}
\usepackage{amsfonts}
\usepackage{amsbsy}
\usepackage{float}
\usepackage{pstricks}
\usepackage{graphicx}
\usepackage{multirow}
\usepackage{hyperref}
\usepackage[]{subfigure}

\begin{document}

\title{\boldmath Characterising exotic matter driving wormholes}
\author{Marco Chianese \inst{1,2} \and Elisabetta Di Grezia  \inst{2} \and Mattia Manfredonia \inst{1,2} \and Gennaro Miele\inst{1,2}}

\institute{ Dipartimento di Fisica Ettore Pancini, Universit\` a di Napoli Federico II, Complesso
Univ. di Monte S. Angelo, I-80126 Napoli, Italy.  \and INFN, Sezione di Napoli, Complesso Univ. Monte S. Angelo, I-80126 Napoli, Italy \\ \email{chianese@na.infn.it, digrezia@na.infn.it,  manfredonia@na.infn.it, miele@na.infn.it}}

\date{Received: date / Revised version: date}
\abstract{ 
In this paper, we develop an iterative approach to span the whole set of exotic matter models able to drive a traversable wormhole. The method, based on a Taylor expansion of metric and stress-energy tensor components in a neighbourhood of the wormhole throat, reduces the Einstein equation to an infinite  set of algebraic conditions, which can be satisfied order by order. The approach easily allows the implementation of further conditions linking the stress-energy tensor components among each  other, like symmetry conditions or equations of state. The method is then applied to some relevant examples of exotic matter characterised by a  constant energy density and that also show an isotropic behaviour in the stress-energy tensor or obeying to a quintessence-like equation of state.}

\maketitle

\section{Introduction}

Wormhole solutions to Einstein equation, from the topological point of view,  can be seen as handles connecting two asymptotically flat regions of the space-time manifold, i.e. a short-cut or a bridge linking together two distant regions of the same space-time. Such solutions were conceived as particle models by Einstein himself  in 1935 (Einstein-Rosen bridge  \cite{Einstein:1935tc,Flamm})\footnote{A predecessor of wormholes is Flamm paraboloid \cite{Flamm}.}. However, an Einstein-Rosen solution turns out to be just a portion of the Schwarzschild's metric \cite{wald,Kruskal:1959vx} describing a Black Hole (BH), therefore the bridge cannot be crossed \cite{Visser,Fuller:1962zza}. In 1988, the seminal work by K. Thorne and M. Morris \cite{Morris:1988cz,Morris:1988tu} led to a more deep understanding of wormhole physics, in particular introducing the concept of a {\it traversable} wormhole. The authors in Ref.  \cite{Morris:1988cz} showed that in order to achieve traversability one has to demand the metric to be event horizons free\footnote{ Though a wormhole does not have an event horizon, and differs in several other important features from a BH, it is hardly distinguishable from it from an observational point of view \cite{Visser,Morris:1988cz,Hawking_large,Tsukamoto:2012xs,Cardoso:2016rao}. Note that a way to identify a wormhole is provided by the peculiar gravitational lensing  of light rays passing near it 
\cite{Bambi:2013nla,Cramer:1994qj,Torres:1998xd,Takahashi:2013jqa,Abe:2010ap,Tsukamoto:2012zz,Tsukamoto:2016zdu,Tsukamoto:2017edq}.}. This feature, combined with Birkhoff's theorem, implies that a traversable asymptotically flat wormhole must be a solution of Einstein equation in presence of some kind of matter (so-called ``{\it exotic matter}'') with non standard stress-energy tensor, namely in a neighbourhood of the  wormhole throat the radial tension $\tau$ must exceed the total energy density $\rho$.  The region around the throat, where such matter is confined \cite{Morris:1988cz,Lobo:2007zb,Barcelo:1999hq}, shows a violation of the Null Energy Condition (NEC) \cite{Visser,Morris:1988tu,Krasnikov:1999ie,Tipler:1978zz}. Note that the relation $\tau > \rho$ implies that an observer radially moving into a wormhole with velocity close to the speed of light observes negative values for the mass-energy density associated with the exotic matter.

In order to study possible sources of a traversable wormholes, one typically postulates some equation of state for the non vanishing components of stress-energy tensor of exotic matter and then tests if it is compatible with the wormhole metric 
\cite{Morris:1988cz,Lobo:2005us,Lobo:2006mt,Lobo:2009du,Lobo:2006ue}. In this work we follow a different approach in order to characterise, on general grounds, the whole set of exotic matter models able to drive 
wormhole metrics. The result is based on the expansion of  the metric and the stress-tensor components in a neighbourhood of the wormhole throat that allows to transform the Einstein equation in an infinite set of algebraic conditions. Interestingly, with an appropriate choice of free parameters, the algebraic equations, up to some order in the expansion, admit a simple solution that can be {\it iteratively} generalised to all orders. The method allows to implement, in a straightforward way, possible further conditions linking the stress-energy tensor components among each other, i.e symmetry conditions or equations of state. 

The paper is organised as follows. In Section 2 we briefly describe a traversable static spherically-symmetric wormhole and the conditions that must be imposed on the corresponding source candidates. In Section 3 we propose our iterative method able to characterise exotic matter close to the wormhole throat. In Section 4 we apply the method for relevant examples of exotic models. Finally we give our conclusions in Section 5.

\section{Traversable Wormholes and Exotic Matter}

Let us consider a static and spherically symmetric traversable wormhole. According to Ref. \cite{Morris:1988cz} the most general metric without horizons in the Schwarzschild coordinate system $(t,r,\theta ,\phi)$~\cite{Morris:1988cz,Visser,Visser:1992qh,Lobo:2016zle} is
\begin{equation}
ds^2 = g_{\mu\nu}dx^{\mu}dx^{\nu} = -e^{2\Phi(r)} dt^2 +\left(1-\frac{b(r)}{r}\right)^{-1} dr^2+r^2\left(d\theta^2+\sin^2\theta  d\phi^2\right),
\label{eq:w_metric}
\end{equation}
where the quantities $b(r)$ and $\Phi(r)$ denote the so-called {\it shape} and {\it red-shift} functions, respectively.\\ Although the previous metric does not have any horizon, further constraints have to be imposed in order to obtain a traversable wormhole (see Ref.~\cite{Morris:1988cz} for a detailed discussion).  In fact one has to satisfy the {\it flaring outward condition} implying~\cite{Morris:1988cz,Visser}
\begin{equation}
b'(r) < \frac{b(r)}{r}\,,
\label{eq:foc}
\end{equation}
near the wormhole ``{\it throat}'', which is defined as the narrowest region of the wormhole where the radial coordinate takes its minimum value, hereafter denoted by $r_0$~\cite{Morris:1988cz,Visser}. Furthermore, it can be shown that $r$ and $b(r)$ must have the same value at the throat,  namely $b(r_0)=r_0$~\cite{Morris:1988cz}. Hence in  $r_0$ one gets from Eq.~(\ref{eq:foc})
\begin{equation}
b'(r_0) < 1 \,.
\label{eq:bminorone}
\end{equation}

According to Birkhoff's theorem a traversable wormhole solution to Einstein equation, showing the desired symmetry properties, can only be achieved in the presence of matter possessing a non-trivial stress-energy tensor. In order to identify such class of peculiar stress-energy tensors it is convenient to adopt the  {\it proper reference frame} $(\hat{t},\hat{r},\hat{\theta},\hat{\phi})$, which is defined as the coordinate system at rest in the Schwarzschild frame $(t,r,\theta ,\phi)$. According to Ref.~\cite{Morris:1988cz}, the transformations between these two frames are given by
\begin{eqnarray}
\mathbf{e}_{\hat{t}} =e^{-\Phi}\mathbf{e}_t \,,\,\,\,\,\,\, \mathbf{e}_{\hat{r}}=\sqrt{1-\frac{b}{r}}\mathbf{e}_r\,,\,\,\,\,\,\, \mathbf{e}_{\hat{ \theta}}=\frac{\mathbf{e}_\theta}{r}\,,\,\,\,\,\,\, \mathbf{e}_{\hat{\phi}}=(r \sin \theta)^{-1}\mathbf{e}_\phi \,.
\label{hatted_transf}
\end{eqnarray}
In this system of coordinates, the metric $g_{\mu \nu}$ locally reduces to the Minkowski's one. Remarkably, in the proper reference frame, the Einstein tensor $G_{\mu \nu}$ results to be diagonal and its non-vanishing components are~\cite{Morris:1988cz}
\begin{eqnarray}
G_{\hat{t} \hat{t}}&=& \frac{b'}{r^2} \,, \nonumber \\
G_{\hat{r} \hat{r}}&=& -\frac{b}{r^3}+2\left(1-\frac{b}{r} \right) \frac{\Phi'}{r} \,, \nonumber \\
G_{\hat{\theta} \hat{\theta}}&=&G_{\hat{\phi} \hat{\phi}}= \left( 1- \frac{b}{r} \right)   \left[ \Phi'' -\frac{b'r-b}{2r(r-b)}\Phi'+(\Phi')^2+\frac{\Phi'}{r}-\frac{b'r-b}{2r^2(r-b)}\right]. 
   \label{tens:Einstein}
\end{eqnarray}
The Einstein equation implies that the stress-energy tensor $T_{\mu \nu}$ has to be diagonal as well, hence one can parametrise it as follows
\begin{equation}
\begin{array}{lllll}
T_{\hat{t} \hat{t}}=\rho (r) \,, & \qquad & T_{\hat{r} \hat{r}}=-\tau (r) \,, & \qquad & T_{\hat{\theta} \hat{\theta}}=T_{\hat{\phi} \hat{\phi}}= p(r) \,,
\end{array}
\label{eq:Tmunu}
\end{equation}
where the quantities $\rho (r)$, $\tau (r)$ and $p(r)$ denote the total mass-energy density, the tension per unit area in the radial direction (the opposite of radial pressure), and the pressure in lateral directions (the transverse pressure), respectively. The flaring outward condition reported in Eq.~(\ref{eq:foc}) can be rewritten in terms of the $T_{\mu \nu}$ components. In particular, near the wormhole's throat (where $r$ and $b(r)$ approaches the same value $r_0$) the tension $\tau (r)$ must exceed the energy density $\rho (r)$
\begin{equation} 
\tau(r) > \rho(r) \,.
\label{eq:flaring}
\end{equation}
 It turns out that in order to get a traversable wormhole the matter has to exhibit a very ``{\it exotic}'' behaviour. Any material satisfying such property (\ref{eq:flaring}) is defined {\it exotic}. It violates the Null Energy Condition (NEC)~\cite{Visser,Morris:1988cz,Visser:1999de,Lobo:2016zle}. The flaring outward condition, provided by the two equivalent expressions in Eq.~(\ref{eq:bminorone})~and~(\ref{eq:flaring}), plays a relevant role in the identification of viable sources for wormhole metric as it will be clarified in the following.

\section{Equation of State for Exotic Matter in a Neighbourhood of the  Wormhole Throat}

As shown in the previous Section, in the proper reference frame the expressions of $G_{\mu \nu}$ and $T_{\mu \nu}$ tensors simplify, hence the Einstein equation
\begin{equation}
G_{\mu \nu} =\frac{1}{m_P^2} \, T_{\mu \nu}\, \,,
\end{equation} 
where $m_P$ stands for the reduced Planck mass in natural units, reads
\begin{eqnarray}
&&\frac{b'}{r^2} =  \frac{1}{m_P^2} \, \rho (r) \,, \nonumber \\
&& -\frac{b}{r^3}+2\left(1-\frac{b}{r} \right) \frac{\Phi'}{r} =  - \frac{1}{m_P^2} \tau (r) \,, \nonumber \\
&&\left( 1- \frac{b}{r} \right)   \left[ \Phi'' -\frac{b'r-b}{2r(r-b)}\Phi'+(\Phi')^2+\frac{\Phi'}{r}-\frac{b'r-b}{2r^2(r-b)}\right] = \frac{1}{m_P^2} p(r) \,.
\label{Einstein_eq}
\end{eqnarray}
By using $m_P$ one can recast all quantities involved in Eq. (\ref{Einstein_eq}) in terms of dimensionless ones, namely $\bar{r} \equiv m_P \, r$, $\bar{b} \equiv m_P \, b$, $\bar{\Phi} = \Phi$, $\bar{\rho} \equiv  \rho/m_P^4$, $\bar{\tau} \equiv  \tau/m_P^4$ and $\bar{p} \equiv  p/m_P^4$ hence one gets
\begin{eqnarray}
&&\frac{\bar{b}'}{\bar{r}^2} = \bar{\rho} (\bar{r}) \,, \nonumber \\
&&\frac{\bar{b}}{\bar{r}^3}-2\left(1-\frac{\bar{b}}{\bar{r}} \right) \frac{\bar{\Phi}'}{\bar{r}} = \bar{\tau} (\bar{r}) \,, \nonumber \\
&&\left( 1- \frac{\bar{b}}{\bar{r}} \right)   \left[ \bar{\Phi}'' -\frac{\bar{b}'\bar{r}-\bar{b}}{2 \bar{r}(\bar{r}-\bar{b})}\bar{\Phi}'+(\bar{\Phi}')^2+\frac{\bar{\Phi}'}{\bar{r}}-\frac{\bar{b}'\bar{r}-\bar{b}}{2\bar{r}^2(\bar{r}-\bar{b})}\right] = \bar{p} (\bar{r}) \,,  \label{Einstein_eq_adim}
\end{eqnarray}
where hereafter the {\it prime} ($'$) denotes the derivative with respect to $\bar{r}$.

In the standard approach once provided a particular expression for stress-energy tensor one can use the Einstein equation (\ref{Einstein_eq_adim}) to look for the functions $\bar{b}$ and $\bar{\Phi}$ of the wormhole metric (see for instance Ref.s~\cite{Morris:1988cz,Lobo:2005us,Lobo:2006mt,Lobo:2009du,Lobo:2006ue}). As one may expect, only for particular type of matter a wormhole solution can be found. 

In this paper we assume a different approach. We aim to characterise the whole set of exotic matter models able to drive a wormhole, and to achieve such a goal we expand all quantities entering in Eq.s  (\ref{Einstein_eq_adim}) around the throat, placed at $\bar{r}=\bar{r}_0\equiv r_0 \, m_P$, namely 
\begin{eqnarray}
& \bar{b}(\bar{r}) = \bar{r}_0 \sum_{n=0}^{\infty}\frac{\bar{b}_n}{n!} \left(\frac{\bar{r}-\bar{r}_0}{\bar{r}_0}\right)^n \,, \,\,\,\,\,\,\,\, &
\bar{\Phi}(\bar{r}) = \sum_{n=0}^{\infty} \frac{\bar{\Phi}_n}{n!} \left(\frac{\bar{r}-\bar{r}_0}{\bar{r}_0}\right)^n \,, \nonumber\\
& \bar{\rho}(\bar{r}) = \frac{1}{\bar{r}_0^2} \sum_{n=0}^{\infty} \frac{\bar{\rho}_n}{n!} \left(\frac{\bar{r}-\bar{r}_0}{\bar{r}_0}\right)^n \,, \,\,\,\,\,\,\,\, 
& \bar{\tau}(\bar{r}) = \frac{1}{\bar{r}_0^2} \sum_{n=0}^{\infty} \frac{\bar{\tau}_n}{n!} \left(\frac{\bar{r}-\bar{r}_0}{\bar{r}_0}\right)^n \,,  \nonumber\\
& \bar{p}(\bar{r}) = \frac{1}{\bar{r}_0^2} \sum_{n=0}^{\infty}  \frac{\bar{p}_n}{n!} \left(\frac{\bar{r}-\bar{r}_0}{\bar{r}_0}\right)^n \,. &
\label{eq: expansion_series}
\end{eqnarray}
The dimensionless coefficients appearing in the above expressions are given by
\begin{eqnarray}
\bar{b}_n&=(\bar{r}_0)^{n-1}\left.\left[ \frac{\partial^n \bar{b}(\bar{r})} {\partial \bar{r}^n} \right] \right|_{\bar{r}_0} \,, \,\,\,\,\,\,\,\,    
\bar{\phi}_n&=\left.\left[ \frac{\partial^n \bar{\phi}(\bar{r})} {\partial \bar{r}^n} \right] \right|_{\bar{r}_0} \,,\\
\bar{\rho}_n&=(\bar{r}_0)^{n-2}\left.\left[ \frac{\partial^n \bar{\rho}(\bar{r})} {\partial \bar{r}^n} \right] \right|_{\bar{r}_0} \,, \,\,\,\,\,\,\,\,  
\bar{\tau}_n&=(\bar{r}_0)^{n-2}\left.\left[ \frac{\partial^n \bar{\tau}(\bar{r})} {\partial \bar{r}^n} \right] \right|_{\bar{r}_0} \,, \\
\bar{p}_n&=(\bar{r}_0)^{n-2}\left.\left[ \frac{\partial^n \bar{p}(\bar{r})} {\partial \bar{r}^n} \right] \right|_{\bar{r}_0} \,.
\label{expand}
\end{eqnarray} 
By substituting such expansions in Eq.s (\ref{Einstein_eq_adim}) one obtains, order by order in the relative distance from the throat $(\bar{r}-\bar{r}_0)/\bar{r}_0$, a set of algebraic conditions. Note that in view of the results of previous section one has to fix $\bar{b}_0 =1$ and $\bar{b}_1<1$. Listing the algebraic equations up to the second order in the expansion one gets
\begin{itemize}
\item {\it zero-th-order conditions}
\end{itemize}
\begin{eqnarray}
\bar{\rho}_0 & = & \bar{b}_1 \,, \label{eq:ro0}\\
\bar{\tau}_0 & = & 1 \,, \label{eq:tau0}\\
\bar{p}_0 & = & \frac{1}{2} \left(1 - \bar{b}_1\right)  \left(1+\bar{\Phi}_1 \right) \label{eq:p0}\,.
\end{eqnarray}
\begin{itemize}
\item {\it first-order conditions}
\end{itemize}
\begin{eqnarray}
\bar{\rho}_1 & = &\bar{b}_2-2 \bar{\rho}_0 \,, \label{eq:ro1}\\
\bar{\tau}_1 & = & - 2 \left(1 - \bar{b}_1\right) \bar{\Phi}_1+  \left(\bar{b}_1 - 3 \right) \,, \label{eq:tau1} \\
\bar{p}_1 & = & \frac{1}{2} \left[\left(1 - \bar{b}_1\right)\left( 3\bar{\Phi}_1 + 3 \bar{\Phi}_2 + 2 \bar{\Phi}_1^2 \right) -\bar{b}_2 \left( 1 + \bar{\Phi}_1\right) - 6 \bar{p}_0\right] \label{eq:p1}\,.
\end{eqnarray}
\begin{itemize}
\item {\it second-order conditions}
\end{itemize}
\begin{eqnarray}
\bar{\rho}_2 & = & \bar{b}_3 - 2 \bar{\rho}_0 - 4 \bar{\rho}_1  \,, \label{eq:ro2}\\
\bar{\tau}_2 & = & -4  \left(1 - \bar{b}_1\right) \left( \bar{\Phi}_1 + \bar{\Phi}_2  \right) + \bar{b}_2 \left(1 + 2 \bar{\Phi}_1\right) - 6 \left(  1 +  \bar{\tau}_1\right) \,, \label{eq:tau2}\\
\bar{p}_2 & = & \frac{1}{2} \left[\left(1 - \bar{b}_1\right) \left( 8 \bar{\Phi}_1^2 + 8 \bar{\Phi}_1 \bar{\Phi}_2 + 4 \bar{\Phi}_1 + 14 \bar{\Phi}_2 + 5 \bar{\Phi}_3\right) +  \right. \nonumber \\
&-& \left. \bar{b}_2 \left( 2 \bar{\Phi}_1^2 +5 \bar{\Phi}_1 +4  \bar{\Phi}_2 +1 \right)-\bar{b}_3 \left( 1 + \bar{\Phi}_1 \right) -12 \left( \bar{p}_0 + \bar{p}_1\right)  \right] \,. \label{eq:p2}
\end{eqnarray}
It is worth observing that all the previous equations hold for any value of $\bar{r}_0$. Hence, one can choose an arbitrary value of throat size and then obtains the corresponding dimensional quantities entering in Eq.s (\ref{Einstein_eq}).

As one can see from equations (\ref{eq:ro0})-(\ref{eq:p2}) for each order of the expansion five new coefficients enter into the game, whereas three new conditions must be satisfied. Hence this implies that one can arbitrarily fix two of such quantities to obtain all the others. Since we are interested in characterising the source candidate for a wormhole,  we consider, order by order, as free parameters the values of $\bar{\rho}_i$ and $\bar{p}_i$ (total energy density and transverse pressure). Thus we get from the set of algebraic equations the corresponding tension in the radial direction $\bar{\tau}_i$ and the metric parameters of wormhole $\bar{b}_i$ and $\bar{\Phi}_i$. Applying such method to previous set of equations one gets
\begin{itemize}
\item {\it zero-th-order solutions}
\end{itemize}
\begin{eqnarray}
\bar{b}_1  & = & \bar{\rho}_0 < 1 \,, \label{eq:ro0_s}\\
\bar{\tau}_0 & = & 1 \,, \label{eq:tau0_s}\\
\bar{\Phi}_1 & = &  \frac{2 \bar{p}_0}{1 - \bar{\rho}_0} -1 \label{eq:p0_s}\,.
\end{eqnarray} 
\begin{itemize}
\item {\it first-order solutions}
\end{itemize}
\begin{eqnarray}
\bar{b}_2  & = & 2 \bar{\rho}_0+ \bar{\rho}_1 \,, \label{eq:ro1_s}\\
\bar{\tau}_1 & = & - \left(4  \bar{p}_0  +  \bar{\rho}_0 + 1 \right) \,, \label{eq:tau1_s} \\
\bar{\Phi}_2 & = & \frac{6 \bar{p}_0 + 2 \bar{p}_1}{3\left(1 - \bar{\rho}_0\right)} + \frac{2 \bar{p}_0 \left(2 \bar{\rho}_0+  \bar{\rho}_1\right) }{3 \left(1 - \bar{\rho}_0\right)^2} - \frac 13 \left( \frac{2 \bar{p}_0}{1 - \bar{\rho}_0} -1 \right)  \left( \frac{4 \bar{p}_0}{1 - \bar{\rho}_0}+ 1 \right)
\label{eq:p1-s}\,.
\end{eqnarray}
\begin{itemize}
\item {\it second-order solutions}
\end{itemize}
\begin{eqnarray}
 \bar{b}_3 & = & 2 \bar{\rho}_0 + 4 \bar{\rho}_1 +  \bar{\rho}_2 \,, \label{eq:ro2_s}\\
\bar{\tau}_2 & = & -4  \left(1 - \bar{\rho}_0\right) \left( \bar{\Phi}_1 + \bar{\Phi}_2  \right) + \bar{b}_2 \left(1 + 2 \bar{\Phi}_1\right) - 6 \left(  1 +  \bar{\tau}_1\right) \,, \label{eq:tau2_s}\\
\bar{\Phi}_3 & = &\left[ 2 \bar{p}_2 +\bar{b}_2 \left( 2 \bar{\Phi}_1^2 +5 \bar{\Phi}_1 +4  \bar{\Phi}_2 +1 \right)+\bar{b}_3 \left( 1 + \bar{\Phi}_1 \right) +12 \left( \bar{p}_0 + \bar{p}_1\right) + \right. \nonumber \\
&& - \left. \left(1 - \bar{\rho}_0 \right) \left( 8 \bar{\Phi}_1^2 + 8 \bar{\Phi}_1 \bar{\Phi}_2 + 4 \bar{\Phi}_1 + 14 \bar{\Phi}_2 \right) \right]\frac{1}{5\left(1 - \bar{\rho}_0\right)} \,, \label{eq:p2_s}
\end{eqnarray}
where for seek of brevity we have not substituted in Eq.s (\ref{eq:tau2_s}) and (\ref{eq:p2_s}) the expressions (\ref{eq:p0_s}), (\ref{eq:ro1_s}), (\ref{eq:p1-s}) and (\ref{eq:ro2_s}).

This approach provides an algebraic iterative method applicable up to any order of the expansion in the relative distance from the throat. In other words, one fixes arbitrarily the functions $\bar{\rho}(\bar{r})$ and $ \bar{p}(\bar{r})$, hence getting the corresponding coefficients in the expansions (\ref{expand}), and then solves the algebraic equations, order by order, for the coefficients of $\bar{\tau}(\bar{r})$, $\bar{\Phi}(\bar{r})$ and $\bar{b}(\bar{r})$.  The method, by allowing to fix arbitrarily $\bar{\rho}(\bar{r})$ and $ \bar{p}(\bar{r})$ in order to get the only $\bar{\tau}(\bar{r})$ compatible with a wormhole metric, explicitly provides all viable models for the exotic matter.

The set of equations (\ref{eq:ro0_s})--(\ref{eq:p2_s}) further simplifies if one considers additional constraints linking each other the components of the stress-energy tensor, like for example a generic equation $F(\bar{\rho},\bar{p},\bar{\tau})=0$. As will be clarified in the following, this is the case of symmetry conditions or equations of state. By using the expansions (\ref{expand}) and substituting them into the expression of $F(\bar{\rho},\bar{p},\bar{\tau})$ we get further algebraic conditions that have to be added, order by order, to the previous list of Eq.s~(\ref{eq:ro0_s})--(\ref{eq:p2_s}). In particular, we have
\begin{itemize}
\item {\it additional zero-th-order condition}
\end{itemize}
\begin{equation}
F_0 =  F\left(\bar{\rho}_0,\bar{p}_0,\bar{\tau}_0\right) = 0 \,.
\end{equation}
\begin{itemize}
\item {\it additional first-order condition}
\end{itemize}
\begin{equation}
F_1=\frac{1}{\bar{r}^2_0}\left[\left.\frac{\partial F}{\partial \bar{\rho}}\right |_{\bar{r}_0}\bar{\rho}_1 + 
\left.\frac{\partial F}{\partial \bar{p}}\right |_{\bar{r}_0}\bar{p}_1 + 
\left.\frac{\partial F}{\partial \bar{\tau}}\right |_{\bar{r}_0}\bar{\tau}_1 \right] = 0\,.
\end{equation}
\begin{itemize}
\item {\it additional second-order condition}
\end{itemize}
\begin{eqnarray}
F_2 & = & \frac{1}{\bar{r}^4_0}\left[
\left.\frac{\partial^2 F}{\partial \bar{\rho}^2} \right |_{\bar{r}_0} \bar{\rho}_1^2+
\left.\frac{\partial^2 F}{\partial \bar{p}^2} \right |_{\bar{r}_0} \bar{p}_1^2+
\left.\frac{\partial^2 F}{\partial \bar{\tau}^2} \right |_{\bar{r}_0} \bar{\tau}_1^2+ \right. \nonumber \\
&& + 2 \left(\left.\frac{\partial^2F}{\partial \bar{\rho} \ \partial \bar{p}}\right |_{\bar{r}_0} \bar{\rho}_1 \bar{p}_1 + 
\left.\frac{\partial^2F}{\partial \bar{\rho} \ \partial \bar{\tau}}\right |_{\bar{r}_0} \bar{\rho}_1 \bar{\tau}_1 + 
\left.\frac{\partial F}{\partial \bar{p} \ \partial \bar{\tau}}\right |_{\bar{r}_0}  \bar{p}_1 \bar{\tau}_1\right) + \nonumber \\
&&  + \left. \left.\frac{\partial F}{\partial \bar{\rho}}\right |_{\bar{r}_0}\bar{\rho}_2 + 
\left.\frac{\partial F}{\partial \bar{p}}\right |_{\bar{r}_0}\bar{p}_2 + 
\left.\frac{\partial F}{\partial \bar{\tau}}\right |_{\bar{r}_0}\bar{\tau}_2 \right] = 0 \,.
\end{eqnarray}
The addition of such further equations reduces the d.o.f of the solutions. In particular one cannot fix arbitrarily both $\bar{\rho}_i$ and $\bar{p}_i$ at any order, but just one of the two.

Far away from the wormhole throat, the metric must approach the Schwartzschild's vacuum solution~\cite{Visser,Morris:1988cz}. A common choice to implement such condition is to bound the exotic matter into a finite spherical region of the space-time of radius $\bar{R}$. Such finite region ($\bar{r}\leq \bar{R}$) is then surrounded by a shell of different material~\cite{Visser,Morris:1988cz,gravitation.Thorne}. Outside the shell, the space-time metric must be the spherically symmetric solution to Einstein equation in vacuum (Schwartzschild space-time) \cite{Morris:1988cz,gravitation.Thorne}\footnote{For examples of  possible different outer space-time see \cite{Bozza:2015haa} and references therein.}. Moreover, the size of the finite region containing the exotic material cannot be completely arbitrary. Indeed, one has to require that the shape function $\bar{b}(\bar{r})$ has only one point where $\bar{b}(\bar{r})=\bar{r}$, which by construction is already verified for $\bar{r}=\bar{r}_0$. If a radius $\bar{r}^*>\bar{r}_0$ such that $\bar{b}(\bar{r}^*)=\bar{r}^*$ exists, one has to demand that the exotic material is confined in a smaller region of the space-time, providing the upper bound $\bar{R} < \bar{r}^*$.

Following Ref.~\cite{Morris:1988cz}, we consider a spherically symmetric junction region with radius $\bar{R}$ and thickness (of transition layer) $\Delta \bar{R}$, filled by a material with constant energy density $\bar{\rho}$ and transverse pressure $\bar{p}$, and linear decreasing $\bar{\tau}$. This configuration ensures that in the region $\bar{R} \leq \bar{r} \leq \bar{R}+\Delta \bar{R}$ (transition layer) the functions featuring in Einstein equation read~\cite{Morris:1988cz}

\begin{align}
 &\bar{b}(\bar{r})=\bar{b}(\bar{R})+\left(\frac{\bar{r}-\bar{R}}{\Delta \bar{R}}\right)\bar{b}(\bar{R}) \,, & \,\,\,\,\,\,\,\, 
\bar{\phi}' (\bar{r})&=\left( \frac{\bar{r}-\bar{R}}{\Delta \bar{R}} \right)\frac{\bar{b}(\bar{R})}{\bar{R}^2} \,, \nonumber
\\
&\bar{\rho}(\bar{r})=\left( \frac{\bar{R}}{\Delta \bar{R}}\right) \bar{\tau}(\bar{R})  \,, &\, \,  \, \,\,\,\,\,\,\,\, 
 \bar{\tau}(\bar{r})&=\bar{\tau}(\bar{R}) - \left(\frac{\bar{r}-\bar{R}}{\Delta \bar{R}}\right)\bar{\tau}(\bar{R}) \,,  \nonumber
\\
 &\bar{p}(\bar{r})=\left( \frac{\bar{R}}{2\Delta \bar{R}}\right) \bar{\tau}(\bar{R}) \, .& &   
\label{eq:junction}
\end{align}
In the region outside the shell $(\bar{r}> \bar{R}+\Delta \bar{R})$, the shape function $\bar{b}(\bar{r})$ is constant and its value is given by $\bar{b}(\bar{R}+\Delta \bar{R})=2\bar{b}(\bar{R}) = 2 \bar{M}$, where $\bar{M}$ is the dimensionless mass featuring in Schwartzchild's solution. From the Einstein equation, we have
\begin{equation}
\bar{M} = \int_{\bar{r}_0}^{\bar{R}} \bar{\rho}(\bar{r}) \bar{r}^2 d\bar{r} + \bar{r}_0 \,.
\label{eq:mass}
\end{equation}
Such a mass is in general a function of the quantities involved in the stress-energy tensor of the exotic material driving the wormhole and depends on the size of the wormhole throat $\bar{r}_0$ and on $\bar{R}$ as well.

Finally, the matching between the wormhole solution in the region $\bar{r}<\bar{R}$ and the Schwartzchild's one in the region $\bar{r}>\bar{R}+\Delta\bar{R}$ determines the value taken by the redshift function at the wormhole throat $\left(\bar{\Phi}(\bar{r}_0)=\bar{\Phi}_0\right)$. Such a value is indeed not involved in the Einstein equation as can be seen from Eq.s~(\ref{eq:ro0_s})--(\ref{eq:p2_s}). By considering the junction functions in the thick shell reported in Eq.s~(\ref{eq:junction}), one obtains
\begin{equation}
\bar{\Phi}_0 = \frac12 \ln \left(1-\frac{2\bar{M}}{\bar{R}+\Delta\bar{R}} \right) - \frac{\Delta\bar{R}}{2}\frac{\bar{b}(\bar{R})}{\bar{R}^2} - \sum_{n=1}^{\infty} \frac{\bar{\Phi}_n}{n!} \left(\frac{\bar{R}-\bar{r}_0}{\bar{r}_0}\right)^n \,.
\end{equation}

\section{The Method for Relevant Examples of Exotic Models}

In the present session we present few examples of interesting exotic models that could resemble reasonable physical systems at least in terms of their properties. In all these cases the output of the approach are the proper radial distributions of stress-energy tensors capable to drive a wormhole, provided at a certain order of approximation.

\subsection{Sphere of isotropic exotic material with constant energy density}

An interesting example of the above method is provided by considering a sphere of isotropic ($F (\bar{\rho},\bar{p},\bar{\tau})= \bar{\tau} + \bar{p} = 0$) exotic matter characterised by constant energy density $\bar{\rho}(\bar{r})=\bar{\rho}_0$ up to a radius $\bar{R}$ (see footnote\footnote{A similar case with constant energy density but assigned equation of state (Generalised Chaplying Gas) has been studied in Ref.s~\cite{Lobo:2005vc,Wang:2016fzy,Gorini:2009em,Gorini:2008zj}.}). In this case one gets
\begin{eqnarray}
&\bar{b}(\bar{r})  &= \bar{r}_0\left \{ 1+\frac{\bar{\rho}_0}{3}\left[ \left(\frac{\bar{r}}{\bar{r}_0} \right)^3-1\right ] \right \} \,,\nonumber\\
&\bar{\Phi}(\bar{r})& =   \bar{\Phi}_0 -  \frac{3-\bar{\rho}_0}{1-\bar{\rho}_0} \left(\frac{\bar{r}-\bar{r}_0}{\bar{r}_0}\right) -  \frac{(7-\bar{\rho}_0)(3-\bar{\rho}_0)}{6(1-\bar{\rho}_0)^2} \left(\frac{\bar{r}-\bar{r}_0}{\bar{r}_0}\right)^2+ \nonumber \\ & &- \frac{(3-\bar{\rho}_0)(7-4\bar{\rho}_0+\bar{\rho}^2_0)}{3 (1- \bar{\rho}_0)^3}\left(\frac{\bar{r}-\bar{r}_0}{\bar{r}_0}\right)^3 + {\cal O}\left(\left(\frac{\bar{r}-\bar{r}_0}{\bar{r}_0}\right)^4\right) \,, \nonumber\\
& \bar{\tau}(\bar{r})&  =  \frac{1}{\bar{r}_0^2} \left[ 1 + \left(3 - \bar{\rho}_0 \right) \left(\frac{\bar{r}-\bar{r}_0}{\bar{r}_0}\right) + \frac{2(12-7\bar{\rho}_0+\bar{\rho}^2_0)}{3(1-\bar{\rho}_0)} \left(\frac{\bar{r}-\bar{r}_0}{\bar{r}_0}\right)^2+ {\cal O}\left(\left(\frac{\bar{r}-\bar{r}_0}{\bar{r}_0}\right)^3\right)\right] \,,\nonumber\\
&\bar{p}(\bar{r})&  =   - \bar{\tau}(\bar{r}) \,.
\end{eqnarray}
It is worth observing that the expression of the shape function is exactly determined, namely the algebraic conditions for  $\bar{b}_n$ with $n \geq 4$ do not involve $\bar{\rho}_0$. On the other hand, in case of constant mass-energy density one may also integrate straightforwardly the first of Eq.s (\ref{Einstein_eq_adim}). Note that for $\bar{\rho}_0 \in \, ]0,1[$ there exists a real positive value $\bar{r}^*\neq \bar{r}_0$ such that $\bar{b}(\bar{r}^*)=\bar{r}^*$
\begin{equation} \label{rstar}
\bar{r}^*= \frac{\bar{r}_0}{2} \left(\sqrt{\frac{12}{\bar{\rho}_0}-3} - 1\right) \,.
\end{equation}
Since we want $\bar{r}_0$ to be the only fixed point of $\bar{b}(\bar{r})$, exotic materials with positive $\bar{\rho}_0$ properly drive traversable wormholes only if $\bar{R} < \bar{r}^*$. On the other hand, if one assumes $\bar{\rho}_0<0$, the condition $\bar{r}^*\neq \bar{r}_0$ cannot be satisfied. Finally, by the expressions of the shape function $\bar{b}(\bar{r})$ and the junction functions reported in Eq.s~(\ref{eq:junction}), we obtain the corresponding $\bar{M}$ of the Schwartzchild's solution in case of a constant and isotropic sphere of exotic matter. In particular, according to Eq.~(\ref{eq:mass}) we have
\begin{equation}
\bar{M} = \, \bar{r}_0 \left\{ 1+\frac{\bar{\rho}_0}{3}\left[ \left(\frac{\bar{R}}{\bar{r}_0} \right)^3-1\right ] \right \}  \,.
\label{mass_SC}
\end{equation} 
Using the expression of $\bar{r}^*$ as upper bound of $\bar{R}$, one gets the maximum Schwartzchild mass obtainable for positive constant mass-energy density that reads 
\begin{equation}
\bar{M}_{\rm max} = \, \bar{r}_0 \left\{ 1+\frac{\bar{\rho}_0}{3}\left[ \frac 18 \left(\sqrt{\frac{12}{\bar{\rho}_0}-3} - 1\right)^3-1\right ] \right \}  \quad {\rm for} \quad 0 < \bar{\rho}_0 < 1 \,.
\label{mass_SC_max}
\end{equation} 
Note that varying $\bar{\rho}_0 \in \, ]0,1[$, the mass $\bar{M}_{\rm max}$ results not  bounded from above.

It is worth observing that under the condition of isotropy and assuming this time a constant pressure the flaring outward constraint cannot be satisfied, hence no traversable wormholes can be constructed with such matter.

\subsection{Quintessence-like exotic materials}

Finally, we examine two different quintessence-like exotic materials (see Ref.s~\cite{Lobo:2005us,Lobo:2006mt,Wang:2016fzy,Zaslavskii:2005fs,Sushkov:2005kj}). In the first case, we consider an exotic model characterised by constant mass-energy density $\bar{\rho}(\bar{r})=\bar{\rho}_0$ and by the equation of state
\begin{equation}
F (\bar{\rho},\bar{p},\bar{\tau})= \bar{p}(\bar{r}) - \omega\,\bar{\rho}(\bar{r}) = 0\,.
\end{equation}
From Eq.s~(\ref{eq:ro0_s})--(\ref{eq:p2_s}), we obtain
\begin{eqnarray}
&&\bar{b}(\bar{r})  = \bar{r}_0\left \{ 1+\frac{\bar{\rho}_0}{3}\left[ \left(\frac{\bar{r}}{\bar{r}_0} \right)^3-1\right ] \right \} \, ,\nonumber\\
&&\bar{\Phi}(\bar{r})  =   \bar{\Phi}_0 - \frac{1-\bar{\rho}_0(1+2\omega)}{1-\bar{\rho}_0} \left(\frac{\bar{r}-\bar{r}_0}{\bar{r}_0}\right)+\frac{1+2\bar{\rho}_0(4\omega-1)+\bar{\rho}_0^2(1-4\omega+8\omega^2)}{6(1-\bar{\rho}_0)^2} \left(\frac{\bar{r}-\bar{r}_0}{\bar{r}_0}\right)^2  \nonumber\\
&&+\frac{-9+\bar{\rho}_0(25+22\omega)-\bar{\rho}_0^2(23+12\omega+88\omega^2)+\bar{\rho}_0^3(7+6\omega+48\omega^2+64\omega^3)}{45 (1- \bar{\rho}_0)^3} \left(\frac{\bar{r}-\bar{r}_0}{\bar{r}_0}\right)^3 + {\cal O}\left(\left(\frac{\bar{r}-\bar{r}_0}{\bar{r}_0}\right)^4\right) \,,\nonumber\\
 &&\bar{\tau}(\bar{r})  =  \frac{1}{\bar{r}_0^2} \left\{ 1 - \left[1 + \bar{\rho}_0(1+4\omega) \right] \left(\frac{\bar{r}-\bar{r}_0}{\bar{r}_0}\right)+\right. \nonumber \\&&\left.+ \frac{2 \left[ 2 - \bar{\rho}_0(1-4\omega) - \bar{\rho}_0^2 (1+2\omega-8\omega^2) \right]}{3(1-\bar{\rho}_0)} \left(\frac{\bar{r}-\bar{r}_0}{\bar{r}_0}\right)^2+ {\cal O}\left(\left(\frac{\bar{r}-\bar{r}_0}{\bar{r}_0}\right)^3\right)\right\} \,, \nonumber\\
&&\bar{p}(\bar{r})  =   \omega \, \bar{\rho}_0 \,.
\end{eqnarray}
Since we are assuming constant mass-energy density also in this case, the same considerations of previous example provided in Eq.s~(\ref{rstar}),~(\ref{mass_SC})~and~(\ref{mass_SC_max}) hold.

In the second case, the quintessence-like exotic material is described by two equations of state
\begin{eqnarray}
F (\bar{\rho},\bar{p},\bar{\tau}) & =&  \bar{p}(\bar{r}) - \omega \,\bar{\rho}(\bar{r}) = 0 \,, \\
G (\bar{\rho},\bar{p},\bar{\tau}) & = &  \bar{\tau}(\bar{r}) - \gamma \,\bar{\rho}(\bar{r}) = 0 \,.
\end{eqnarray}
From these two constraints and from the Einstein equation, one gets
\begin{eqnarray}
&\bar{b}(\bar{r}) & = \bar{r}_0\left \{ 1 + \frac{1}{\gamma} \left(\frac{\bar{r}-\bar{r}_0}{\bar{r}_0}\right) + \frac{\gamma - 4\omega - 1}{2 \gamma^2} \left(\frac{\bar{r}-\bar{r}_0}{\bar{r}_0}\right)^2 \right. \nonumber \\
&& \left. + \frac{2\gamma^3 + 16\gamma \left(2 + \omega\right)\left(1+4\omega\right) + 3\left(1+4\omega\right)^2-\gamma^2\left(37+168\omega \right)}{18\left(\gamma -1\right)\gamma^3} \left(\frac{\bar{r}-\bar{r}_0}{\bar{r}_0}\right)^3+ {\cal O}\left(\left(\frac{\bar{r}-\bar{r}_0}{\bar{r}_0}\right)^4\right) \right\}\, ,\nonumber\\
&\bar{\Phi}(\bar{r})&  =  \bar{\Phi}_0 + \frac{2\omega -\gamma +1}{\gamma -1} \left(\frac{\bar{r}-\bar{r}_0}{\bar{r}_0}\right) +\frac{\left(2\omega -\gamma +1\right)\left(1 -\gamma - 8\omega\right)}{6\left(\gamma-1\right)^2} \left(\frac{\bar{r}-\bar{r}_0}{\bar{r}_0}\right)^2  \nonumber\\
&&+\frac{1}{45\left(\gamma-1\right)^3\gamma}\left[\left(1-\gamma\right)^3\left(1+9\gamma\right) + 2 \left(\gamma -1\right)\left[-8 + \gamma \left(-19 + 9\gamma\right)\right]\omega \right. \nonumber \\
&& \left.- 80\left(\gamma - 1\right)\left(1+3\gamma\right)\omega^2 + 64\left(2+3\gamma\right)\omega^3 \right] \left(\frac{\bar{r}-\bar{r}_0}{\bar{r}_0}\right)^3 + {\cal O}\left(\left(\frac{\bar{r}-\bar{r}_0}{\bar{r}_0}\right)^4\right) \,,\nonumber\\
& \bar{\rho}(\bar{r})&  =  \frac{1}{\bar{r}_0^2} \left\{ \frac{1}{\gamma} - \frac{4\omega + \gamma +1}{\gamma^2} \left(\frac{\bar{r}-\bar{r}_0}{\bar{r}_0}\right) \right. \nonumber \\ &&+ \left. \frac{8\gamma^3+3\left(1+4\omega\right)^2+4\gamma\left(1+4\omega\right)\left(5+4\omega\right)-\gamma^2\left(31+120\omega\right)}{6\left(\gamma-1\right)\gamma^3} \left(\frac{\bar{r}-\bar{r}_0}{\bar{r}_0}\right)^2 + {\cal O}\left(\left(\frac{\bar{r}-\bar{r}_0}{\bar{r}_0}\right)^3\right)    \right\} \,, \nonumber\\
&\bar{p}(\bar{r})&  = \omega \, \bar{\rho} (\bar{r}) \,, \nonumber \\
&\bar{\tau}(\bar{r})& =  \gamma \, \bar{\rho} (\bar{r}) \,.
\label{last_case_bf}
\end{eqnarray}
It is worth observing that the flaring outward condition leads to the constraint
\begin{equation}
\bar{b}_1  =  \bar{\rho}_0 < 1 \quad \Rightarrow \quad \gamma > 1 \,.
\end{equation}
Such a constraint also implies that the mass-energy density is positive near the wormhole throat. Substituting the approximated expression for $\bar{\rho}(\bar{r})$ in Eq.~(\ref{eq:mass}), we obtain the mass $\bar{M}$ of the exterior Schwarzschild space-time:
\begin{eqnarray}
&\bar{M}& \simeq \bar{r}_0\left\{1+ \frac{1}{\gamma}\left(\frac{\bar{R}-\bar{r}_0}{\bar{r}_0}\right)+\frac{\gamma-4\omega-1}{2\gamma^2}\left(\frac{\bar{R}-\bar{r}_0}{\bar{r}_0}\right)^2\right.\nonumber \\
&&+\frac{2\gamma^3+16\gamma (2+\omega)(1+4\omega)+3(1+4\omega)^2-\gamma^2(37+168\omega)}{9(\gamma-1)\gamma^3}\left(\frac{\bar{R}-\bar{r}_0}{\bar{r}_0}\right)^3 \nonumber \\\
&&+\frac{5\gamma^2+3(1+4\omega)^2+\gamma(1+4\omega)(23+16\omega)-\gamma^2(31+132\omega)}{12(\gamma-1)\gamma^3}\left(\frac{\bar{R}-\bar{r}_0}{\bar{r}_0}\right)^4\nonumber \\
&&\left. +\frac{8\gamma^2+3(1+4\omega)^2+4\gamma(1+4\omega)(5+4\omega)-\gamma^2(31+120\omega)}{30(\gamma-1)\gamma^3}\left(\frac{\bar{R}-\bar{r}_0}{\bar{r}_0}\right)^5 \right\} \,. 
\end{eqnarray}
As in previous examples, we have to require $\bar{R} \leq \bar{r}^*$, where $\bar{r}^*>\bar{r}_0$ is the solution of equation $\bar{b}(\bar{r}^*)=\bar{r}^*$ (if it exists). In particular, we have

\begin{eqnarray}
&\bar{r}^*&=\frac{1}{D}\left[-5\gamma^3+6(1+4\omega)^2+\gamma(1+4\omega)(55+32\omega)-4\gamma^2(14+75\omega) \right. \nonumber \\
&&\left. +(\gamma-1)\gamma\sqrt{16\gamma^2+33(1+4\omega)^2+2\gamma(1+4\omega)(119+64\omega)-7\gamma^2(41+192\omega)} \right] \,, \nonumber \\
&&
\label{eq:quint_r*}
\end{eqnarray}
where
\begin{equation}
D= 2 \left[ 2\gamma^3+16\gamma(2+\omega)(1+4\omega)+3(1+4\omega)^2-\gamma^2(37+169\omega) \right] \,.
\end{equation}
This implies that $\bar{r}^*$ takes a real value only if, for a given value of $\gamma>1$, the quantity $\omega$ satisfies the following constraint
\begin{equation}
 \left(528+512\gamma \right)\omega^2 + \left(264+1080\gamma-1344\gamma^2\right) \omega + 33+238\gamma-278\gamma^2+16\gamma^3 >0 \,.
\end{equation}
For all the values of $\gamma>1$ and $\omega$ satisfying the above inequality, the radius $\bar{r}^*$ of Eq.~(\ref{eq:quint_r*}) provides the maximum allowed value for $\bar{R}$, implying a corresponding maximum value for the mass $\bar{M}$.

\section{Conclusions}
 
We describe a simple approach able to span the whole set of exotic matter models driving a wormhole solution of Einstein equation. The method is based on a Taylor expansion of the stress-energy tensor components (total energy density,  the radial tension and the transverse pressure) and of the wormhole metric (the shape and the red-shift functions) in the relative distance from the wormhole throat. By substituting such expansions in the Einstein equation one transforms the three nonlinear differential equations in an infinite set of algebraic conditions for the coefficients of the five expansions. For each order of the expansion we have five new coefficients entering in the set of equations and at the same time  three new conditions to be satisfied. This allows to fix arbitrarily two quantities out of five. If one chooses to assign the total energy density and the transverse pressure,  all the other quantities, namely the radial tension, the shape and the red-shift functions are completely determined. Note that the algebraic equations, order by order can be easily solved and this allows for an iteration of the procedure. Moreover, the approach makes straightforward the implementation of extra constraints on the kind of exotic model considered, like a generic condition $F(\bar{\rho},\bar{p},\bar{\tau})=0$ linking the stress-energy tensor components among each other. This is the case if one assumes symmetry conditions  (like for example isotropic exotic matter), and/or a particular equation of state. In this case the d.o.f. of the possible solution reduce, and hence one can arbitrarily fix just one quantity between the total energy density and the transverse pressure, having the other completely determined. Limiting the exotic material to sphere of given radius, the matching conditions with the asymptotic Schwartzchild's solution are also discussed.

The method has been applied to some relevant examples of exotic matter characterised by constant mass-energy density and isotropy, or described by quintessence-like equations of state. In all cases the approach provides, up to a certain order of the expansion in the relative distance from the throat, the radial profile of all quantities involved in the physical system.

\section*{\bf Acknowledgments}

The authors acknowledge support by the Instituto Nazionale di Fisica Nucleare I.S. TAsP and the PRIN 2012 ``Theoretical Astroparticle Physics" of the Italian Ministero dell'Istruzione, Universit\`a e Ricerca.

\bibliographystyle{ieeetr}
\bibliography{Exotic_matter_closer_to-throat}

\begin{thebibliography}{10}

\bibitem{Einstein:1935tc}
A.~Einstein and N.~Rosen, ``{The Particle Problem in the General Theory of
  Relativity},'' {\em Phys. Rev.}, vol.~48, pp.~73--77, 1935.

\bibitem{Flamm}
Flamm, ``Beitr\"{a}ge zur einsteinschen gravitationstheorie,'' {\em Phys Z.17},
  1916.

\bibitem{wald}
R.~M.Wald, {\em General Relativity}.
\newblock The university of Chicago Press, 1984.

\bibitem{Kruskal:1959vx}
M.~D. Kruskal, ``{Maximal extension of Schwarzschild metric},'' {\em Phys.
  Rev.}, vol.~119, pp.~1743--1745, 1960.

\bibitem{Visser}
M.~Visser, {\em Lorentzian Wormholes. From Einstein to Hawking}.
\newblock American Institute of Physics, 1996.

\bibitem{Fuller:1962zza}
R.~W. Fuller and J.~A. Wheeler, ``{Causality and Multiply Connected
  Space-Time},'' {\em Phys. Rev.}, vol.~128, pp.~919--929, 1962.

\bibitem{Morris:1988cz}
M.~S. Morris and K.~S. Thorne, ``{Wormholes in space-time and their use for
  interstellar travel: A tool for teaching general relativity},'' {\em Am. J.
  Phys.}, vol.~56, pp.~395--412, 1988.

\bibitem{Morris:1988tu}
M.~S. Morris, K.~S. Thorne, and U.~Yurtsever, ``{Wormholes, Time Machines, and
  the Weak Energy Condition},'' {\em Phys. Rev. Lett.}, vol.~61,
  pp.~1446--1449, 1988.

\bibitem{Hawking_large}
S.~W. Hawking and G.~F.~R. Ellis, {\em The Large Scale Structure of
  Space-Time}.
\newblock Cambridge University Press, 1973.

\bibitem{Tsukamoto:2012xs}
N.~Tsukamoto, T.~Harada, and K.~Yajima, ``{Can we distinguish between black
  holes and wormholes by their Einstein ring systems?},'' {\em Phys. Rev.},
  vol.~D86, p.~104062, 2012.

\bibitem{Cardoso:2016rao}
V.~Cardoso, E.~Franzin, and P.~Pani, ``{Is the gravitational-wave ringdown a
  probe of the event horizon?},'' {\em Phys. Rev. Lett.}, vol.~116, no.~17,
  p.~171101, 2016.
\newblock [Erratum: Phys. Rev. Lett.117,no.8,089902(2016)].

\bibitem{Bambi:2013nla}
C.~Bambi, ``{Can the supermassive objects at the centers of galaxies be
  traversable wormholes? The first test of strong gravity for mm/sub-mm very
  long baseline interferometry facilities},'' {\em Phys. Rev.}, vol.~D87,
  p.~107501, 2013.

\bibitem{Cramer:1994qj}
J.~G. Cramer, R.~L. Forward, M.~S. Morris, M.~Visser, G.~Benford, and G.~A.
  Landis, ``{Natural wormholes as gravitational lenses},'' {\em Phys. Rev.},
  vol.~D51, pp.~3117--3120, 1995.

\bibitem{Torres:1998xd}
D.~F. Torres, G.~E. Romero, and L.~A. Anchordoqui, ``{Might some gamma-ray
  bursts be an observable signature of natural wormholes?},'' {\em Phys. Rev.},
  vol.~D58, p.~123001, 1998.

\bibitem{Takahashi:2013jqa}
R.~Takahashi and H.~Asada, ``{Observational Upper Bound on the Cosmic
  Abundances of Negative-mass Compact Objects and Ellis Wormholes from the
  Sloan Digital Sky Survey Quasar Lens Search},'' {\em Astrophys. J.},
  vol.~768, p.~L16, 2013.

\bibitem{Abe:2010ap}
F.~Abe, ``{Gravitational Microlensing by the Ellis Wormhole},'' {\em Astrophys.
  J.}, vol.~725, pp.~787--793, 2010.

\bibitem{Tsukamoto:2012zz}
N.~Tsukamoto and T.~Harada, ``{Signed magnification sums for general spherical
  lenses},'' {\em Phys. Rev.}, vol.~D87, no.~2, p.~024024, 2013.

\bibitem{Tsukamoto:2016zdu}
N.~Tsukamoto and T.~Harada, ``{Light curves of light rays passing through a
  wormhole},'' {\em Phys. Rev.}, vol.~D95, no.~2, p.~024030, 2017.

\bibitem{Tsukamoto:2017edq}
N.~Tsukamoto, ``{Retrolensing by a wormhole: $\pi$ and $3\pi$ in the sky?},''
  2017.

\bibitem{Lobo:2007zb}
F.~S.~N. Lobo, ``{Exotic solutions in General Relativity: Traversable wormholes
  and 'warp drive' spacetimes},'' 2007.

\bibitem{Barcelo:1999hq}
C.~Barcelo and M.~Visser, ``{Traversable wormholes from massless conformally
  coupled scalar fields},'' {\em Phys. Lett.}, vol.~B466, pp.~127--134, 1999.

\bibitem{Krasnikov:1999ie}
S.~Krasnikov, ``{A Traversable wormhole},'' {\em Phys. Rev.}, vol.~D62,
  p.~084028, 2000.
\newblock [Erratum: Phys. Rev.D76,109902(2007)].

\bibitem{Tipler:1978zz}
F.~J. Tipler, ``{Energy conditions and spacetime singularities},'' {\em Phys.
  Rev.}, vol.~D17, pp.~2521--2528, 1978.

\bibitem{Lobo:2005us}
F.~S.~N. Lobo, ``{Phantom energy traversable wormholes},'' {\em Phys. Rev.},
  vol.~D71, p.~084011, 2005.

\bibitem{Lobo:2006mt}
F.~S.~N. Lobo, ``{Traversable wormholes supported by cosmic accelerated
  expanding equations of state},'' in {\em {Recent developments in theoretical
  and experimental general relativity, gravitation and relativistic field
  theories. Proceedings, 11th Marcel Grossmann Meeting, MG11, Berlin, Germany,
  July 23-29, 2006. Pt. A-C}}, pp.~2193--2195, 2006.

\bibitem{Lobo:2009du}
F.~S.~N. Lobo and J.~P. Mimoso, ``{Possibility of hyperbolic tunneling},'' {\em
  Phys. Rev.}, vol.~D82, p.~044034, 2010.

\bibitem{Lobo:2006ue}
F.~S.~N. Lobo, ``{Van der Waals quintessence stars},'' {\em Phys. Rev.},
  vol.~D75, p.~024023, 2007.

\bibitem{Visser:1992qh}
M.~Visser, ``{Dirty black holes: Thermodynamics and horizon structure},'' {\em
  Phys. Rev.}, vol.~D46, pp.~2445--2451, 1992.

\bibitem{Lobo:2016zle}
F.~S.~N. Lobo, ``{From the Flamm–Einstein–Rosen bridge to the modern
  renaissance of traversable wormholes},'' {\em Int. J. Mod. Phys.}, vol.~D25,
  no.~07, p.~1630017, 2016.

\bibitem{Visser:1999de}
M.~Visser and C.~Barcelo, ``{Energy conditions and their cosmological
  implications},'' in {\em {Proceedings, 3rd International Conference on
  Particle Physics and the Early Universe (COSMO 1999): Trieste, Italy,
  September 27-October 3, 1999}}, pp.~98--112, 2000.

\bibitem{gravitation.Thorne}
C.~Misner, K.~Thorne, and J.~Wheeler, {\em Gravitation}.
\newblock W. H. Freeman, 1973.

\bibitem{Bozza:2015haa}
V.~Bozza and A.~Postiglione, ``{Alternatives to Schwarzschild in the weak field
  limit of General Relativity},'' {\em JCAP}, vol.~1506, no.~06, p.~036, 2015.

\bibitem{Lobo:2005vc}
F.~S.~N. Lobo, ``{Chaplygin traversable wormholes},'' {\em Phys. Rev.},
  vol.~D73, p.~064028, 2006.

\bibitem{Wang:2016fzy}
D.~Wang and X.-H. Meng, ``{Geometric dark energy traversable wormholes
  constrained by astrophysical observations},'' 2016.

\bibitem{Gorini:2009em}
V.~Gorini, A.~{\relax Yu}. Kamenshchik, U.~Moschella, O.~F. Piattella, and
  A.~A. Starobinsky, ``{More about the Tolman-Oppenheimer-Volkoff equations for
  the generalized Chaplygin gas},'' {\em Phys. Rev.}, vol.~D80, p.~104038,
  2009.

\bibitem{Gorini:2008zj}
V.~Gorini, U.~Moschella, A.~{\relax Yu}. Kamenshchik, V.~Pasquier, and A.~A.
  Starobinsky, ``{Tolman-Oppenheimer-Volkoff equations in presence of the
  Chaplygin gas: stars and wormhole-like solutions},'' {\em Phys. Rev.},
  vol.~D78, p.~064064, 2008.

\bibitem{Zaslavskii:2005fs}
O.~B. Zaslavskii, ``{Exactly solvable model of wormhole supported by phantom
  energy},'' {\em Phys. Rev.}, vol.~D72, p.~061303, 2005.

\bibitem{Sushkov:2005kj}
S.~V. Sushkov, ``{Wormholes supported by a phantom energy},'' {\em Phys. Rev.},
  vol.~D71, p.~043520, 2005.

\end{thebibliography}

\end{document}